\documentclass{jpp}
\usepackage{graphicx}

\usepackage[utf8]{inputenc}
\usepackage[T1]{fontenc}
\usepackage[draft]{pgf}
\usepackage{amsmath}
\usepackage{graphicx}
\usepackage{dcolumn}
\usepackage{bm}
\usepackage{natbib}
\usepackage{soul}
\usepackage{xcolor}
\usepackage{amsmath}
\usepackage{amssymb}

\usepackage{color}

\usepackage{subfigure}
\usepackage[T1]{fontenc}
\usepackage{enumerate}
\usepackage{makecell}
\usepackage{multirow}
\usepackage{hyperref}
\usepackage{array}

\shorttitle{Drift kinetic theory of alpha particle transport by NTMs}
\shortauthor{E. A. Tolman and P.J. Catto}

\title{Quasilinear drift kinetic theory of alpha particle transport by neoclassical tearing modes}

\author{Elizabeth A. Tolman\aff{1,2,3}  
  \corresp{\email{tolman@ias.edu}}
 \and Peter J.  Catto\aff{1} }

\affiliation{\aff{1}Plasma Science and Fusion Center, Massachusetts Institute of Technology,
Cambridge, MA 02139, USA \aff{2}Institute for Advanced Study,
Princeton, NJ 08540, USA \aff{3}Center for Computational Astrophysics, Flatiron Institute, 162 Fifth Avenue, New York, NY, 10010, USA
 }

\begin{document}

\maketitle

\begin{abstract}
Kinetic theory of particles near resonances is a current topic of discussion in plasma physics and astrophysics. We extend this discussion to the kinetic theory of the interaction between alpha particles (energetic particles predicted to exist in large quantities in next-generation fusion experiments) and a neoclassical tearing mode (NTM), a resistively-driven perturbation which sometimes exists in a tokamak. We develop a quasilinear treatment of the interaction between alphas and an NTM, showing why an NTM can be a source of significant passing alpha particle transport in tokamaks.  The limitations on quasilinear theory constrain our theory's applicability to small amplitude NTMs, highlighting the importance of nonlinear studies.
\end{abstract}

\section{Introduction}
Many recent studies have considered kinetic theory of particles near resonances, with application to both plasma physics \citep{white2018resonances,catto2018ripple,white2019collisional,tolman2019dependence,duarte2019collisional,catto2019collisional,catto2020collisional,lestz2021analytic,catto2021lower,catto2023merging,white2023assessment,duarte2023shifting} and galactic dynamics \citep{binney2020angle,chiba2022oscillating,hamilton2022galactic}. These works are interesting from a fundamental perspective: the structure of kinetic theory has intrinsic intellectual value. Furthermore, the works have practical value. In plasma physics, resonant interactions between magnetic and electric field perturbations and energetic particles can lead to transport that reduces fusion device performance and damages device walls \citep{kurki2009ascot,scott2020fast}.  In galactic dynamics, resonant interactions help to determine a galaxy's structure \citep{dehnen2000effect} and influence the results of cosmological galaxy formation simulations \citep{roshan2021fast,hamilton2022galactic}.

In the recent literature, a type of resonance that remains less explored is one that exists in a tokamak between passing energetic particles and the neoclassical tearing mode (NTM). The NTM is a special type of resistive tearing mode which results from the bootstrap current created naturally by the toroidal nature of tokamaks \citep{la2006neoclassical}. Experimentally, NTMs are seen to produce significant energetic particle transport \citep{mynick1993transport,garcia2007ntm,heidbrink2018phase}. Wave-particle resonance with the NTM differs from resonance with other mode classes studied in recent papers: the NTM has a very small frequency \citep{buttery2003onset,hirvijoki2014ascot} and is characterized by toroidal mode number $n$ and poloidal mode number $m$ related by $nq = m$, with $q$ the tokamak safety factor at the center of the mode; both $n$ and $m$ are typically small.

In this paper, we aim to fill a gap in the literature. In particular, we derive the quasilinear theory of passing alpha particle transport by NTMs. Very close to the resonant surface that exists at the center of the NTM, the mode resonates with energetic particles with vanishing tangential drift frequency, which are freely passing. Slightly away from the resonant surface, the mode-particle resonance can move closer to the pitch angle of the trapped-passing boundary. Resonance causes alpha particle transport which can deplete the slowing down distribution at even low mode amplitudes.

The paper begins in Section~\ref{sec:setup} with a discussion of the tokamak equilibrium and the NTM perturbation that affects it. Then, in Section~\ref{sec:quaform}, we discuss the quasilinear approximation which is used to study alpha particle transport, and use it to derive an expression for the alpha particle energy flux.  In Section~\ref{sec:response}, we derive the perturbed alpha particle distribution necessary to evaluate the energy flux, and discuss the nature of the resonance between the energetic particles and the mode. In Section~\ref{sec:flux}, we evaluate the particle flux that results from this resonance, and use it to estimate the mode amplitude at which depletion of the slowing down distribution would occur.  In Section~\ref{sec:concl}, we summarize our results, discuss their implications for tokamak plasma physics, and describe the limits on the validity of our treatment. 

\section{Equilibrium and NTM perturbation}
\label{sec:setup} 
In this Section, we first describe the tokamak magnetic equilibrium in which the alpha particle transport occurs. Then, we describe the NTM perturbation that modifies this equilibrium. Finally, we present the parameters that describe alpha particles in this equilibrium. 
\subsection{Description of equilibrium}
The coordinates describing the tokamak equilibrium are $\psi$ (the poloidal flux function), $\vartheta$ (the poloidal angle), and $\alpha$, defined by
\begin{equation}
\label{eq:alphadef}
\alpha \equiv \zeta - q\left(\psi\right) \vartheta.
\end{equation}
Here, $\zeta$ is the toroidal angle, with $\left| \nabla \zeta \right| = R^{-1}$, where $R$ is the major radius coordinate, and $q\left(\psi\right)$ is the safety factor.
We study the behavior of alpha particles confined by a magnetic field which is stationary and axisymmetric except for the NTM perturbation. Such a field is stated in the Clebsch representation \citep{kruskal1958equilibrium,grad1958hydromagnetic,d1991clebsch} as
\begin{equation}
\label{eq:equilibb}
\vec{B} = \nabla \alpha \times \nabla \psi =I \left( \psi \right) \nabla \zeta + \nabla \zeta \times \nabla \psi,
\end{equation}
with $I\left(\psi \right)$ characterizing the strength of the toroidal magnetic field by $B_\zeta = I\left(\psi\right)/R$. We do not consider background electric fields because they do not affect alpha particle trajectories as strongly as the magnetic fields do (a discussion is found in Section 2 of \cite{tolman2020drift}). The unit vector corresponding to the magnetic field is
\begin{equation}
\frac{\vec{B}}{B} \equiv \hat{b}.
\end{equation}
The poloidal angle is chosen such that\footnote{Later in the paper, an approximate form of this definition,  $q \left(\psi \right) \approx  \left(R \hat{b} \cdot  \nabla \vartheta  \right)^{-1}$, will be used to simplify expressions.}  
\begin{equation} 
\label{eq:sfdef}
\vec{B}\cdot \nabla \vartheta = \left|I \right|/\left[q\left(\psi \right)R^2 \right] = q \left(\psi \right)^{-1} \left|\vec{B} \cdot \nabla \zeta \right|.
\end{equation} 
At this point, $\vartheta$ is a fully general straightened field line poloidal coordinate which must only be chosen such that~\eqref{eq:sfdef} holds.  This allows for shaping and finite aspect ratio. Later, a circular-cross-section,  high-aspect-ratio approximation is used.

The poloidal component of the field is denoted $B_p$, and can be found from $B_p \approx \epsilon B/q$, with the inverse aspect ratio
\begin{equation}
\label{eq:eps}
\epsilon \approx r /R,
\end{equation}
where $r$ is the local minor radius. The flux coordinate and the minor radius are related by
\begin{equation}
\label{eq:psi}
\partial/\partial \psi = \left(1/ RB_p \right) \partial /\partial r.
\end{equation} The shear of the field is given by  
\begin{equation}
\label{eq:shear}
    s\equiv \left(r/q\right) \partial q /\partial r.
\end{equation}
The strength of the on-axis field is given by $B_0$.

The total magnetic field affecting the alpha particles includes both the equilibrium magnetic field and the magnetic field resulting from the NTM, which is  discussed in Section~\ref{sec:perts}. The total magnetic field from all of these sources is denoted $\vec{B}_{\rm tot}$ (with unit vector $\hat{b}_{\rm tot}$). Several parameters can be defined in terms of the total field or the unperturbed field; the total quantities are in general only used before the perturbation analysis is carried out. 

When numerical examples are needed in this paper, we use the equilibrium parameters given in Table~\ref{tab:equilib}, which are similar to those planned for SPARC, a DT tokamak experiment currently being constructed \citep{pablo2020}.

\begin{table}
  \begin{center}
\def~{\hphantom{0}}
  \begin{tabular}{lc}
Parameter & Value \\[3pt] 
$B_{0}$&$1.2 \times 10^{5}$ G \\ 
$R$  & $185$ cm\\
$n_e$, $n_i$  & $4 \times 10^{14}$ cm$^{-3}$\\
$T_e$  & $20$  keV\\
$v_0$  & $1.3 \times 10^{9}$  cm s$^{-1}$\\
  \end{tabular}
  \caption{Example tokamak parameters used in this paper, similar to those planned for SPARC \citep{creely2020,pablo2020}. The bulk plasma is assumed to be an equal mix of deuterium and tritium. For convenience, this Table includes the alpha particle birth speed, $v_0$, even though this parameter is the same in any tokamak.}
  \label{tab:equilib}
  \end{center}
\end{table}
\subsection{Description of neoclassical tearing mode perturbation}
\label{sec:perts}
We consider situations where the equilibrium discussed previously is perturbed by an NTM, which we describe in this Section. At rational surfaces $\psi = \psi_s$ in the tokamak, the safety factor is given by 
\begin{equation}
q\left( \psi = \psi_s \right) = \frac{m}{n}.
\end{equation}
An NTM is a resistively-driven instability at this surface that introduces a perturbation to the vector potential \citep{hirvijoki2012alfven,hirvijoki2014ascot} given approximately by 
\begin{equation}
\label{eq:adef}
\vec{A_1} =  A_\parallel \left(  \psi , \vartheta, \zeta\right) \hat{b},
\end{equation}
where 
\begin{equation}
\label{eq:apar}
    A_\parallel \left(  \psi , \vartheta, \zeta\right) =A_{\parallel} \left(\psi\right) \cos{\left(n\zeta - m \vartheta \right)} =  \Re \left[ A_{\parallel} \left(\psi \right) e^{i n \alpha + i \left(q n - m\right) \vartheta}\right].
\end{equation}
We consider only cases where a single NTM is present.
This mode will be resonant with the magnetic field at the rational surface. At this surface, the perturbation will not vary at all along the direction of the magnetic field; its wavevector will be perpendicular to the magnetic field. The values of $n$ and $m$ are typically low, as higher mode number modes bend field lines more and thus are more stable \citep{buttery2000neoclassical}. A brief explanation of how this mode arises is given in Appendix~\ref{sec:ntm}.
 
The perturbed magnetic field from~\eqref{eq:adef} is given by
\begin{equation}
\label{eq:b1}
\vec{B}_1 = \nabla \times \vec{A}_1 .
\end{equation}
The overall magnetic field unit vector $\hat{b}_{\rm tot}$ is related to the perturbation and the background field by
\begin{equation}
\hat{b}_{\rm tot} \approx \hat{b} + B^{-1} \nabla \times \left( B^{-1} A_\parallel \vec{B} \right) \approx \hat{b} + B^{-1} \nabla \left(I A_\parallel/B \right) \times \nabla \zeta,
\end{equation}
where the last form preserves axisymmetry and perturbs the flux by 
\begin{equation}
\psi_1 = -I A_\parallel/B,
\end{equation}
assuming $B_p \ll B$.

Later, it will be helpful to have expressions for some quantities involving the perturbed magnetic field. For example, we can state the convenient result, obtained using~\eqref{eq:equilibb}, that 
\begin{equation}
\label{eq:identity}
\vec{B}_1 \cdot \nabla \psi = \nabla \cdot \left( \frac{A_\parallel}{B} \vec{B} \times \nabla \psi \right) = \vec{B} \cdot \nabla \left(I A_\parallel/B  \right) - B \partial A_\parallel/\partial\zeta.
\end{equation}
Then, we also have that 
\begin{equation}
\label{eq:relation}
\vec{B}_1 \cdot \nabla \psi + \vec{B} \cdot \nabla \psi_1 = \vec{B} \cdot \nabla \left( \psi_1 + I A_\parallel/B \right) - B \partial_\zeta A_\parallel = -B \partial_\zeta A_\parallel,
\end{equation}
which	vanishes	upon	toroidal averaging,	implying	that	we	can	define	perturbed,	axisymmetric	flux	surfaces.

\subsection{Description of particles}
We study the behavior of a population of alpha particles in tokamak magnetic fields perturbed by an NTM of form~\eqref{eq:adef}. In this Section, we present the system of quasilinear equations used to study this population.  Alpha particles are characterized by their speed, $v$ (or, equivalently, energy normalized to mass, $\mathcal{E} \equiv v^2/2$), the sign of $v_\parallel$, the component of their velocity parallel to the equilibrium magnetic field, and their magnetic moment.
The sign of $v_\parallel$ is given by the variable $\sigma$, defined as follows:
\begin{equation}
\label{eq:sigmadef}
    \sigma = \left\{
     \begin{array}{@{}l@{\thinspace}l}
      0, \, \text{trapped particles}\\
      \frac{v_\parallel}{\left|v_\parallel\right|}, \, \text{passing particles} \\
     \end{array}
  \right. .
\end{equation}
The magnetic moment can be defined relative to the total magnetic field,
\begin{equation}
\label{eq:lambdatot}
\mu_{\rm tot} \equiv \frac{v_{\perp,\rm tot}^2}{2 B_{\rm tot}},
\end{equation}  
or relative to the unperturbed field,
\begin{equation}
\label{eq:lambdal}
\mu \equiv \frac{v_\perp^2}{2B}.
\end{equation} 
(Here,  $v_{\perp,tot}$ is the component of their velocity perpendicular to total field and $v_\perp$ is the component of their velocity perpendicular to the unperturbed field.) Note that these definitions of magnetic moment exclude the alpha particle mass. We can also define
\begin{equation}
    \lambda \equiv 2 \mu B_0/v^2.
\end{equation}

The alpha particle mass and charge number are given by $M_\alpha$  and $Z_\alpha $, respectively; $\Omega_{\rm tot}  \equiv Z_\alpha eB_{\rm tot}/M_\alpha c$ is the gyrofrequency in the total field and $\Omega \equiv Z_\alpha eB/M_\alpha c$ the gyrofrequency in the unperturbed field.  The alpha particle poloidal gyrofrequency (in the unperturbed field) is $\Omega_p \equiv Z_\alpha eB_p/M_\alpha c$. The alpha particle gyroradius is $\rho_\alpha \equiv v_\perp/ \Omega$, and the poloidal gyroradius $\rho_{p \alpha} \equiv v_\perp/ \Omega_p$.

\section{Development of quasilinear formulation}
\label{sec:quaform}

Outside the narrow tearing layer, the NTM perturbation~\eqref{eq:adef} has a radial scale length which is large compared to the alpha gyroradius.  Therefore, we can begin our study from the  drift-kinetic equation \citep{hazeltine1973recursive}:
\begin{multline}
\label{eq:dke}
\frac{\partial f}{\partial t} + \left(v_\parallel \hat{b}_{\rm tot} + \vec{v}_{d, \rm tot} \right) \cdot \nabla f + \left[\frac{Z_\alpha e}{M_\alpha } \left(v_\parallel \hat{b}_{\rm tot} + \vec{v}_{d, \rm tot} \right)\cdot \vec{E}_{\rm tot} + \mu_{\rm tot} \frac{\partial B_{\rm tot}}{\partial t} \right] \frac{\partial f}{\partial \mathcal{E}} 
\\
= C\left\{f\right\} +\frac{S_{\rm fus} \delta \left(v-v_0\right)}{4 \pi v^2}.
\end{multline}
Note that this expression neglects the finite spread of alpha particle birth speeds. Here, the collision operator appropriate for alpha particles with speed much faster than the ion thermal speed is given by \citep{cordey1976effects,catto2018ripple}
\begin{equation}
\label{eq:collop}
C\left\{f \right\} = \frac{1}{\tau_s v^2} \frac{\partial}{\partial v} \left[ \left(v^3 + v_c^3 \right) f \right] + \frac{2 v_\lambda^3 B_{0}}{\tau_s v^3 B}   \frac{v_\parallel}{v} \frac{\partial}{\partial \lambda} \left( \lambda \frac{v_\parallel}{v} \frac{\partial f}{\partial \lambda} \right).
\end{equation}

The first term represents electron and ion drag while the second represents pitch angle scattering off of bulk ions. 
Here, the alpha slowing down time is given by
\begin{equation}
 \tau_s\left(\psi \right) = \frac{3M_\alpha T_e^{3/2}\left(\psi \right)  }{4 \left( 2 \pi m_e \right)^{1/2} Z_\alpha ^2 e^4 n_e\left(\psi \right) \ln \Lambda_c},
 \end{equation}
 with $\ln \Lambda_c$, $T_e$, $n_e$, and $m_e$ the Coulomb logarithm, the electron temperature, the electron density, and the electron mass, respectively.  The critical speed at which alpha particles switch from mainly losing energy to electrons to  mainly losing energy to ions is found by summing over background ions,
 \begin{equation}
 v_c^3\left(\psi \right) = \frac{3 \pi^{1/2} T_e^{3/2} \left(\psi \right)}{\left(2m_e\right)^{1/2} n_e \left(\psi \right)} \sum_i \frac{Z_i^2 n_i\left(\psi \right)}{M_i},
 \end{equation}   
 with $Z_i$, $n_i$, and $M_i$ the charge, density, and mass of each of the background species. This is of similar size to $v_\lambda$, the speed at which pitch angle scattering is important to the behavior of the equilibrium energetic alpha population:
 \begin{equation}
 v_\lambda^3\left(\psi \right) \equiv \frac{3 \pi^{1/2} T_e^{3/2} \left(\psi \right)}{\left(2m_e\right)^{1/2} n_e \left(\psi \right) M_\alpha } \sum_i Z_i^2 n_i\left(\psi \right).
 \end{equation}
The velocity coordinate contains a component along the magnetic field $\hat{b}_{\rm tot}$, $v_{\parallel,\rm tot}$, and a component perpendicular to it, $\vec{v}_{d,\rm tot}$. The drift velocity is given by
\begin{equation}
\label{eq:vdtot}
\vec{v}_{d,\rm tot} =  \frac{\lambda_{\rm tot} v^2}{2 B_0\Omega_{\rm tot}}\hat{b}_{\rm tot} \times \nabla B_{\rm tot} + \frac{v_{\parallel, \rm tot}^2}{\Omega_{\rm tot}} \hat{b}_{\rm tot} \times \left( \hat{b}_{\rm tot} \cdot \nabla \hat{b}_{\rm tot} \right) \approx \Omega_{\rm tot}^{-1} v_{\parallel,\rm tot} \nabla_{\perp, \rm tot} \times \left(v_{\parallel,\rm tot} \hat{b}_{\rm tot} \right),
\end{equation}
where $\nabla_{\perp,\rm tot} = - \hat{b}_{\rm tot} \times \left(\hat{b}_{\rm tot} \times \nabla \right)$. Later, we will use $v_\parallel$ and $\vec{v}_d$ to refer to the parallel velocity and drift in the presence of only the unperturbed field.

For a mode of form~\eqref{eq:adef}, there is no significant perturbed electric field as any time variation of the perturbed magnetic field is unimportant.  In addition, the effect of the tokamak's unperturbed electric field is unimportant relative to the effect of the magnetic drift.  \footnote{For an explanation, see footnote 4 of \cite{tolman2020drift}.} Thus, the coefficient of $\partial f/\partial \mathcal{E}$ can be set to zero. In addition, we neglect the time derivative as we are seeking steady-state solutions in a slowly time varying NTM field. Then, our kinetic equation becomes 
\begin{equation}
\label{eq:initke}
\left(v_{\parallel ,\rm tot}  \hat{b}_{\rm tot} + \vec{v}_{d,\rm tot} \right) \cdot \nabla f 
\\
= C\left\{f\right\} +\frac{S_{\rm fus} \delta \left(v-v_0\right)}{4 \pi v^2}.
\end{equation}
Our goal is to develop approximate expressions from and solutions of~\eqref{eq:initke} that allow us to evaluate the radial energy flux of the passing alpha particles under the influence of the NTM. We begin by recognizing that 
\begin{equation}
    \left(v_{\parallel ,\rm  tot}  \hat{b}_{\rm tot} + \vec{v}_{d,\rm tot} \right) \cdot \nabla f  = \frac{v_{\parallel, \rm tot}}{B_{\rm tot}} \nabla \cdot \left[ f \left(\vec{B}_{\rm tot} + \frac{B_{\rm tot}}{v_{\parallel,\rm tot}} \vec{v}_{d,\rm tot} \right)    \right].
\end{equation}
 We can average~\eqref{eq:initke} over $\zeta$ and over a transit using 
 $d\tau_{\rm tot} = d \vartheta/\left(v_{\parallel,\rm tot} \hat{b}_{\rm tot} \cdot \nabla \vartheta \right)$:
\begin{multline}
  \frac{1 }{2 \pi\oint d\tau_{\rm tot}  }\oint \frac{d\tau_{\rm tot} d \zeta v_{\parallel,\rm tot}}{B_{\rm tot}} \nabla \cdot \left[ f \left(\vec{B}_{\rm tot} + \frac{B_{\rm tot}}{v_{\parallel, \rm tot}} \vec{v}_{d,\rm tot} \right)    \right]
\\
= \frac{1 }{2 \pi \oint d\tau_{\rm tot} } \oint d\tau_{\rm tot} d \zeta 
 \left[C\left\{f\right\} +\frac{S_{\rm fus} \delta \left(v-v_0\right)}{4 \pi v^2}\right].
\end{multline}
Using the divergence in $\psi_{\rm tot} = \psi + \psi_1, \, \vartheta, \, \zeta$ variables, we can simplify the left hand side of the previous equation:
\begin{multline}
       \frac{1 }{2 \pi\oint d\tau_{\rm tot} }\oint \frac{d\tau_{\rm tot} d \zeta v_{\parallel, \rm tot}}{B_{\rm tot}} \nabla \cdot \left[ f \left(\vec{B}_{\rm tot} + \frac{B_{\rm tot}}{v_{\parallel, \rm tot}} \vec{v}_{d,\rm tot} \right)    \right]
     \\
     = \frac{1 }{2 \pi\oint d\tau_{\rm tot} } \frac{\partial}{\partial \psi_{\rm tot}}  \oint \frac{\sigma d\vartheta d \zeta }{\vec{B}_{\rm tot} \cdot \nabla \vartheta } \left[ f \left(\vec{B}_{\rm tot} + \frac{B_{\rm tot}}{v_{\parallel, \rm tot}} \vec{v}_{d,\rm tot} \right) \cdot \nabla \psi_{\rm tot}   \right].
\end{multline}
In this expression, the term $\vec{v}_{d,\rm tot} \cdot \nabla  \psi_{\rm tot}$ drives neoclassical transport, which we do not evaluate here.\footnote{The neglected contribution to the transport can be found as explained in \cite{hsu1990neoclassical} and \cite{catto2018ripple} and is unaffected by the NTM.} Let us now consider the term $f \vec{B}_{\rm tot} \cdot \nabla \psi_{\rm tot}$. First, consider that in the absence of an NTM, the distribution function must be the axisymmetric
\begin{equation}
    f\left(\psi,\vartheta, v, \sigma, \lambda \right) = f_0 \left(\psi, \vartheta, v, \sigma, \lambda\right).
\end{equation}
The introduction of the NTM causes two effects in the distribution function. First, the equilibrium distribution function acquires the structure of the perturbed flux surfaces, i.e., $ f_0 \left(\psi, \vartheta, v, \sigma, \lambda\right) \rightarrow  f_0 \left(\psi + \psi_1, \vartheta, v, \sigma, \lambda \right)$. Second, the distribution function itself may be changed due to the motions resulting from the NTM, i.e., $f_0 \left(\psi + \psi_1, \vartheta, v, \sigma, \lambda \right) \rightarrow f_0 \left(\psi + \psi_1, \vartheta, v, \sigma, \lambda \right) + f_1 \left(\psi + \psi_1, \vartheta, \zeta, v, \sigma, \lambda \right) $. That is, we have
\begin{equation}
    f  =f_0 \left(\psi + \psi_1, \vartheta, v, \sigma, \lambda \right) + f_1 \left(\psi + \psi_1, \vartheta, \zeta, v, \sigma, \lambda \right).
\end{equation}
With this expression, we can evaluate $f \vec{B}_{\rm tot} \cdot \nabla \psi_{\rm tot}$. Neglecting the very small $\zeta$ dependence in $\vec{B}_{\rm tot} \cdot \nabla \psi_{\rm tot}$,  we note that all first order terms go to zero when averaged over $\zeta$, and we can write   
\begin{multline}
\frac{1 }{2 \pi\oint d\tau_{\rm tot} } \frac{\partial}{\partial \psi_{\rm tot}}  \oint \frac{\sigma d\vartheta d \zeta }{\vec{B}_{\rm tot} \cdot \nabla \vartheta }  f \vec{B}_{\rm tot} \cdot \nabla \psi_{\rm tot}  = 
\\ 
\frac{1 }{2 \pi\oint d\tau_{\rm tot} } \frac{\partial}{\partial \psi_{\rm tot}}  \oint \frac{\sigma d\vartheta d \zeta }{\vec{B}_{\rm tot} \cdot \nabla \vartheta } f \left( \vec{B} \cdot \nabla \psi_1 + \vec{B}_1 \cdot \nabla \psi \right). 
\end{multline}
Making use of~\eqref{eq:relation} we find that 
\begin{equation}
\label{eq:qeq}
      \frac{S_{\rm fus} \delta \left(v -v_0\right)\oint_\zeta d\tau  }{4 \pi v^2}  +   \oint_\zeta d\tau C\left\{f_0 \right\} 
    \\
    = - \frac{\partial}{\partial \psi}  \oint \frac{ \sigma d\vartheta d \zeta }{\hat{b} \cdot \nabla \vartheta }  f_1 \partial_\zeta A_\parallel.  
\end{equation}
The subscript $\zeta$ indicates that the integral is performed at a fixed value of $\zeta$.  
When the right hand side of~\eqref{eq:qeq} is negligible, the value of $f_0$ is found to be the slowing down distribution,
\begin{equation}
\label{eq:slowdn}
f_0 \left(\psi, v\right) = \frac{S_{\rm fus}\left(\psi \right) \tau_s\left(\psi \right) H\left(v_0-v\right)}{4 \pi \left[v^3 + v_c^3\left(\psi\right)\right]},
\end{equation} 
with $H$ a unit step function.
To find the energy flux $\Gamma_q$ that results from the right hand side of~\eqref{eq:qeq}, we can multiply by $M_\alpha v^2/2$ and $v_\parallel d^3 v/B$, where (summing over both signs of $v_\parallel$)
\begin{equation} 
\label{eq:d3v}
d^3 v = \frac{2 \pi  B v^3 dv d\lambda}{  B_0 \left|v_\parallel \right| },
\end{equation}
and divide by $\oint_\zeta d\vartheta/\left(\vec{B} \cdot \nabla \vartheta  \right)$. We find, assuming large aspect ratio,
\begin{multline}
\label{eq:heatflux}
\frac{M_\alpha v_0^2}{2}S_{\rm fus} + \frac{M_\alpha }{2 \oint_\zeta 
 d \vartheta /\vec{B}\cdot \nabla \vartheta}  \oint_{\zeta} \frac{d \vartheta }{\vec{B} \cdot \nabla \vartheta } \int d^3 v v^2 C \left\{f_0 \right\} 
 \\
 = \frac{-M_\alpha /2 }{2 \pi \oint d\vartheta / \left(\vec{B}\cdot \nabla \vartheta  \right)} \partial_\psi \left[ \oint \frac{d\vartheta d \zeta }{\vec{B} \cdot \nabla \vartheta } \int d^3v  v^2 v_\parallel  f_1 \partial_\zeta A_\parallel\right]  
 \equiv \frac{1}{r} \frac{\partial}{\partial r } \left( r \Gamma_q \right).
\end{multline}
Then, we can write that the alpha particle energy flux is given by 
\begin{equation}
\label{eq:hflux}
\Gamma_q \approx -\frac{M_\alpha }{4 \pi R B_p \oint d\vartheta / \vec{B}\cdot \nabla \vartheta} \oint \frac{d \vartheta d \zeta }{\vec{B}\cdot \nabla \vartheta} \int d^3v v^2 v_\parallel f_1 \partial_\zeta A_\parallel.
\end{equation}

In the next Section we will find a solution for $f_1$. If unperturbed trajectories are sufficient to calculate the distribution response, it is possible to neglect the nonlinear terms in the kinetic equation.\footnote{The validity of this approximation is discussed in Section~\ref{sec:concl}.}  Then, we need only solve the linear equation
\begin{equation}
\label{eq:lineq}
    \left( v_\parallel \hat{b} +  \vec{v}_d \right)\cdot \nabla f_1 -  v_\parallel \left( \partial_\zeta A_\parallel \right) \left( \partial f_0/\partial \psi \right)
    = C \left\{ f_1 \right\}.
\end{equation}

\section{Plasma response to perturbations}
\label{sec:response}

In this Section, we calculate the plasma response to the NTM perturbation described in the previous Section. Our expression~\eqref{eq:adef} for the vector potential and the use of unperturbed trajectories in~\eqref{eq:lineq} suggests using drift kinetic angular momentum $\psi_\star = \psi - I v_\parallel/\Omega$, $\vartheta$, and $\alpha_\star = \zeta - q_\star \vartheta$ as the  variables, with $q_\star = q \left(\psi_\star\right)$, since $n\zeta - m\vartheta = n\alpha -\left(m - nq \right) \vartheta = n\alpha_\star - \left(m -q_\star n \right)\vartheta$. (Note that these variables mean that to the requisite order $\left(v_\parallel \hat{b} + \vec{v}_d \right) \cdot \nabla \psi_\star = 0$.) Then, we can posit that 
\begin{equation}
\label{eq:fourierf1}
    f_1 
    = \Re {\left[-i F_1 \left(\psi_\star, \vartheta, v , \mu, \sigma \right) e^{i n \alpha_\star   }\right]}.
\end{equation}
Using~\eqref{eq:apar} and~\eqref{eq:lineq}, we find that the equation governing $F_1$ is 
\begin{equation}
\label{eq:govf}
i v_\parallel \hat{b}\cdot \nabla \vartheta  
 \left. \partial_\vartheta F_1 \right|_{\alpha_\star} + i n \omega_\alpha F_1 - e^{inq_\star \vartheta} C \left\{F_1 e^{-in q_\star\vartheta }\right\} = -n v_\parallel \left(\frac{\partial f_0}{\partial \psi} \right) A_{\parallel} e^{-i \left(m - q_\star n \right) \vartheta}.
\end{equation}
Here, we have defined
\begin{equation}
    \omega_{\alpha} \equiv \left(v_\parallel \hat{b} + \vec{v}_d \right) \cdot \nabla \alpha_\star \approx  v_\parallel \hat{b} \cdot \nabla \vartheta \frac{\partial}{\partial \psi} \left( \frac{q I v_\parallel}{\Omega} \right)  .
\end{equation}
(Due to the dominance of parallel streaming over drifts, we have that $v_\parallel \hat{b} \cdot \nabla \vartheta \gg \vec{v}_d \cdot \nabla \vartheta$.)
Let us integrate~\eqref{eq:govf} over the particle trajectory, assuming $A_\parallel \left(\psi \right) \approx A_\parallel \left( \psi_\star \right)$ because $\psi_\star-\psi = - \left( I v_\parallel /\Omega \right) \left(\partial q/\partial \psi \right)$ is assumed small, i.e.,
\begin{multline}
\label{eq:int1}
    \oint d \tau \left[i v_\parallel \hat{b}\cdot \nabla \vartheta  
 \left. \partial_\vartheta F_1 \right|_{\alpha_\star} + i n \omega_\alpha F_1 - e^{inq_\star \vartheta} C \left\{F_1 e^{-in q_\star\vartheta }\right\} \right] =
 \\
 -n 
  A_{\parallel}  \left(\psi_\star\right) \left(\frac{\partial f_0}{\partial \psi} \right) \oint 
 v_\parallel e^{-i \left(m - q_\star n \right) \vartheta} d\tau .
\end{multline}
 We note the radial ponderomotive force departure during a poloidal transit due to the nonlinear radial drift from the NTM drive is small compared to a poloidal gyroradius for our small NTM amplitude ordering as is seen by expanding $A_\parallel \left(\psi+\psi_1\right) \approx A_\parallel-(I/B)dA_\parallel^2/d\psi $.

For trapped particles, whose trajectories traverse a closed path in $\vartheta$, the right side of~\eqref{eq:int1} vanishes, which shows that the trapped particle response vanishes. The drift caused by the perturbation is proportional to the particle parallel velocity, so particles on closed trajectories along which parallel velocity reverses sign experience no significant drift.

\begin{table}
  \begin{center}
\def~{\hphantom{0}}
  \begin{tabular}{lcc}
Quantity & Value  & Description \\ 
$\omega_d \left(v\right)$  & $\frac{nv^2}{\Omega_p R^2}$ & Tangential drift frequency \\ 
$\nu_{\rm pas}\left(v\right) $  &  $\frac{v_\lambda^3}{v^3 \tau_s \epsilon}$ & Pitch angle scattering frequency \\
\\ 
$V \left( v \right)$  &  $\frac{ v q n A_{\parallel}}{\sqrt{\epsilon} R B_0}$ & Drift due to NTM drive \\
  \end{tabular}
  \caption{Values defined in order to simplify expressions in the paper.}
  \label{tab:quants}
  \end{center}
\end{table}

\begin{table}
  \begin{center}
\def~{\hphantom{0}}
  \begin{tabular}{lcc}
Quantity & Value for passing particles & Approximate value used in calculations \\ 
$n \overline{\omega}_{q}  \tau_f$  & $\frac{2 \omega_d \left(v \right) R qk \left[2E \left(k \right) - \left(2-k^2\right)K\left(k\right)  \right]}{v \left[\left(1-\epsilon\right)k^2 + 2 \epsilon \right]\sqrt{2\epsilon \lambda}}$ & $\frac{\omega_d \left(v \right) Rq \log{\left[1-k \right]}}{v\sqrt{2\epsilon}}$\\ 
$\oint d\tau v_\parallel^2$  &  $ \frac{4q  R v \sqrt{2 \epsilon \lambda }}{k} E \left(k \right)$ & $ 4q  R v \sqrt{2 \epsilon} $  \\
\\ 
  \end{tabular}
  \caption{Quantities appearing in the kinetic equation. The complete elliptic integral of the first kind is denoted by $K\left(k \right)$; the complete integral of the second kind is $E\left(k\right)$. (This Table uses quantities defined in Table~\ref{tab:quants}.)}
  \label{tab:res}
  \end{center}
\end{table}
For passing particles there is a significant response when there is a resonance. For the passing particles, we can also Fourier transform in poloidal angle, writing that 
\begin{equation}
\label{eq:fourierf1v2}
    f_1 
    = \Re {\left[-i F_1 \left(\psi_\star, v , \mu, \sigma \right) e^{i n \alpha_\star - i\left( m - q_\star n \right)\vartheta  }\right]}.
\end{equation}
With this new expression, we can again integrate over the passing trajectory, and recognize that $q \approx q_{\star}$ for the collision term and that the magnetic shear term in the drift term will cancel one in the streaming term. We can define a drift without shear,
\begin{equation}
q v_\parallel \hat{b} \cdot \nabla \vartheta \frac{\partial}{\partial \psi} \left( \frac{I v_\parallel}{\Omega}\right) \equiv \omega_q,
\end{equation}
and write that
\begin{equation}
\label{eq:keqa}
    2 \pi \sigma  i \left(nq -m \right) F_1 + in \bar{\omega}_{q} \tau_f F_{1} - \oint d \tau C \left\{F_{1}\right\} = -\frac{2\pi q R  \sigma V \left(v \right)}{v \sqrt{\epsilon}} \frac{\partial f_0}{\partial r},
\end{equation}
where 
\begin{equation}
    \tau_f = \oint d \tau \approx \frac{4qR \sqrt{\left(1-\epsilon\right) k^2 + 2 \epsilon} K \left(k \right)}{v \sqrt{2\epsilon}} \xrightarrow[k\rightarrow 1]{} - \left( \frac{2qR}{v \sqrt{2\epsilon}}\right) \log{\left(1-k \right)},
\end{equation}
\begin{equation}
 \overline{\omega}_q \tau_f \equiv \oint d\tau \omega_q,
\end{equation}
and we have used quantities defined in Table~\ref{tab:quants}. We have adopted the following approximate expression for the strength of the axisymmetric magnetic field:
\begin{equation}
\label{eq:bstrength}
B = B_{0} \left[1- \epsilon\left(\psi \right) \cos \vartheta \right],
\end{equation}
with $\epsilon \approx r /R  \ll 1 $ (as introduced in Section~\ref{sec:setup}).  In this field, the parallel velocity is given by $v_\parallel = \pm v \sqrt{\left[1 - \left(1-\epsilon\right) \lambda \right] - 2 \epsilon \lambda \sin^2{\vartheta/2}}$.
We also introduce the parameter $k$, which is defined in terms of $\lambda$, 
\begin{equation}
\label{eq:kdef}
k^{-2} \equiv \frac{1-\left(1-\epsilon\right)\lambda}{2 \epsilon \lambda}.
\end{equation}
A value of $k= 1$ corresponds to a barely passing particle and a value of $k= 0$ corresponds to a freely passing particle.  [A pedagogical introduction to this parameter can be found in~\citet{helander2005collisional}.]

At a sharp resonance, the pitch angle scattering part of the collision operator~\eqref{eq:collop} is most important. The drift resonance is strongest for the barely passing ($k\rightarrow 1$) alphas because the transit time $\tau_f$ in $\bar{\omega}_q\tau_f$ diverges logarithmically. We can set $\lambda \approx 1$ in the pitch angle scattering part of the collision operator. (This simplification reflects that the most important $\lambda$-dependence in the collision operator comes from the second derivative with respect to $\lambda$ of $F_1$.) Then we can write, using quantities in Table~\ref{tab:quants} and Table~\ref{tab:res},
\begin{equation}
\label{eq:collop2}
 \oint d \tau C \left\{F_{1}\right\}
    \approx \frac{8\sqrt{2\epsilon} \epsilon \nu_{\rm pas}\left(v \right)  qR }{v} \frac{\partial^2 F_1}{\partial \lambda^2} \approx \frac{\nu_{\rm pas}\left(v \right) qR}{\sqrt{2 \epsilon} v } \frac{\partial^2 F_1 }{\partial k^2}.
\end{equation}
Here, we have also modified the derivative in the collision operator with the approximate expression $\partial / \partial \lambda \approx  \left[ 1/\left( 4 \epsilon \right)  \right] \partial / \partial k$.

With the definitions of $\lambda$,~\eqref{eq:lambdal}, and $k$,~\eqref{eq:kdef}, we can evaluate the bounce time, the transit averaged drift, and an average present in the collision operator, which are given in Table~\ref{tab:res}. 
Then, our kinetic equation~\eqref{eq:keqa} becomes 
\begin{equation}
\omega_d \left(v \right) \log{\left[\gamma \left(1-k \right) \right]} F_1 + i \nu_{\rm pas}\left(v \right) \frac{\partial^2 F_1}{\partial k^2} = i 2\pi \sqrt{2} \sigma V \left(v \right)  \frac{\partial f_0}{\partial r},
\end{equation}
where we have defined the normally large factor
\begin{equation}
    \gamma \left(v \right) \equiv e^{ \frac{ 2 \pi \sqrt{2\epsilon}  \sigma v \left(qn-m \right) }{ \omega_d \left(v \right)q R}},
\end{equation}
which increases with distance in flux space from the center of the NTM for $\sigma\left(qn-m\right)>0$. 

\begin{figure*}
\centering
\subfigure{{\includegraphics[width=8cm]{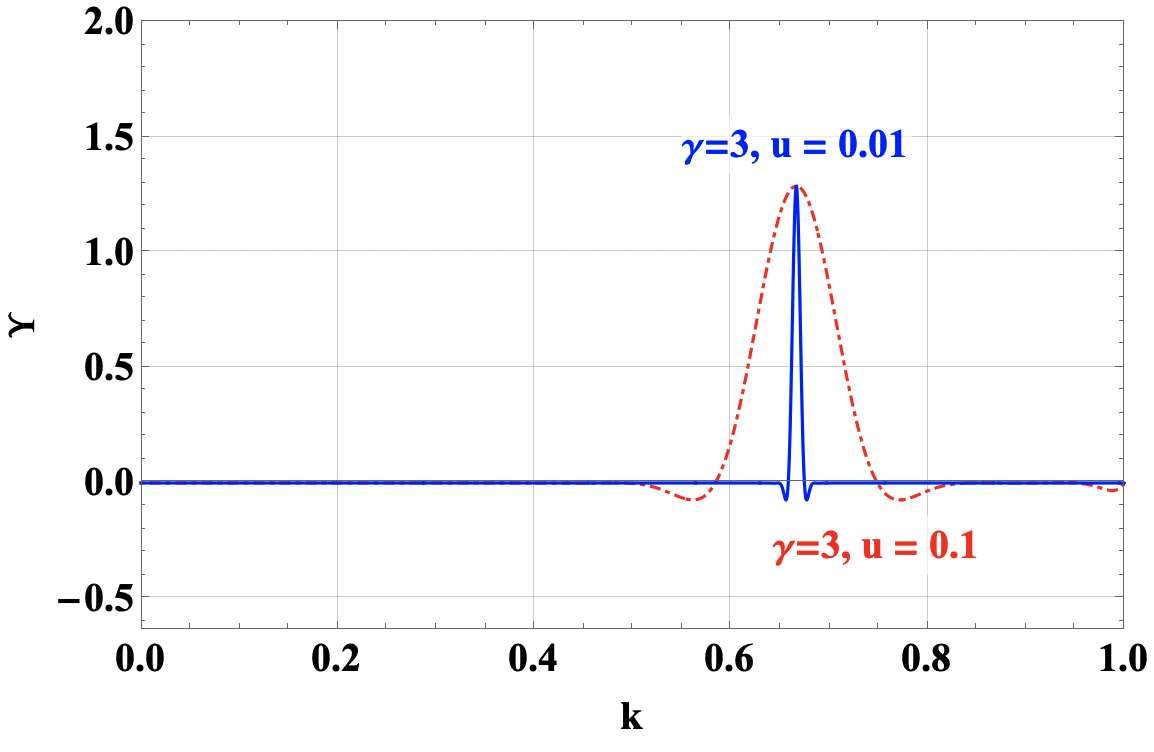} }}
\subfigure{{\includegraphics[width=8cm]{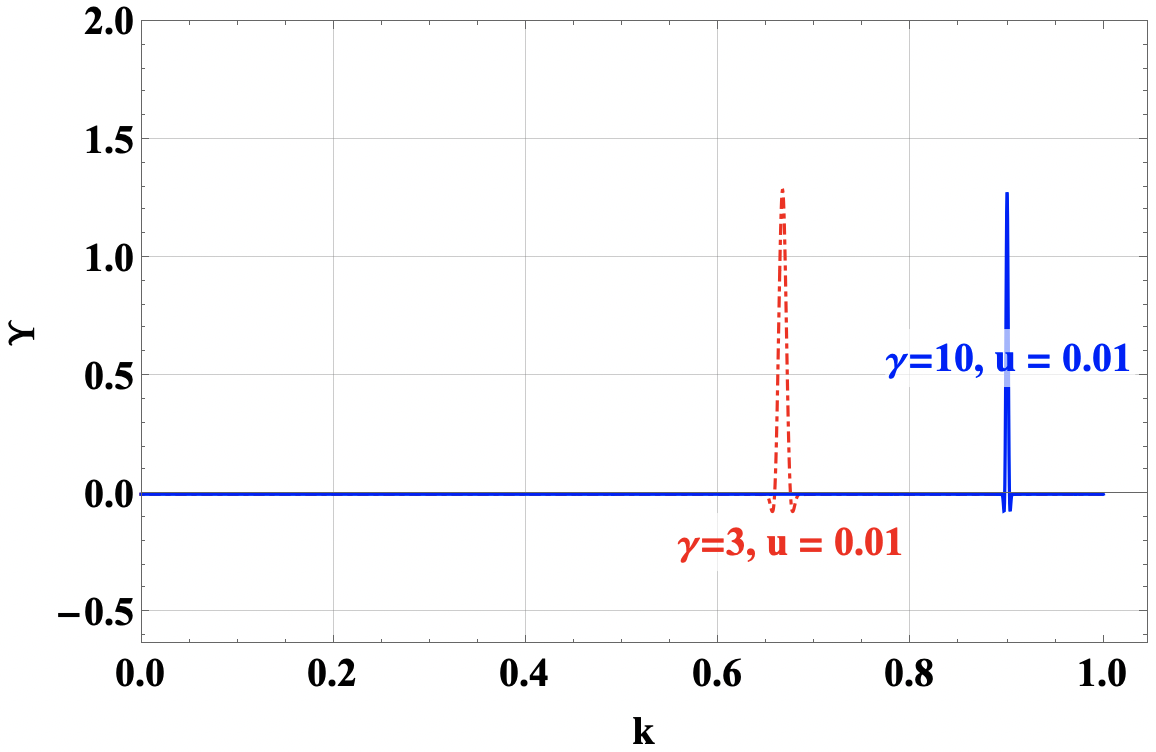} }}
  \caption{Diagrams of plasma response function $\Upsilon$,~\eqref{eq:upsilon}, showing that $\gamma$ controls the proximity of the resonance to the trapped-passing boundary and $u$ controls the width of the resonance response.}
\label{fig:resresponse}
\end{figure*}

In order to simplify the kinetic equation, let us make an additional set of definitions and approximations. First, we define a variable $x$  such that 
\begin{equation}
\label{eq:xdef}
 x  \equiv \gamma \left(1 - k \right),
\end{equation}
which represents distance from the trapped-passing boundary and accounts for the distance from the rational surface of interest. We note that $\gamma \gg 1$ and that expanding around $x=1$ gives
\begin{equation}
\log{x} \approx x -1.
\end{equation}
We also define a small quantity
\begin{equation}
\label{eq:udef}
    u \left(v \right) \equiv \frac{ \gamma^{2/3} \left( v \right)\nu_{\rm pas}^{1/3} \left(v \right)}{\omega_{d}^{1/3} \left(v \right)},
\end{equation}
which represents the width of the resonance, allowing the definition of an order-unity quantity
\begin{equation}
    z \equiv \frac{x-1}{u}.
\end{equation}
Furthermore, we can define a normalized version of the distribution function $\Upsilon\left(z \right)$ such that 
\begin{equation}
\label{eq:F1}
     F_1 \left(z \right)  \equiv  -\left[\frac{2 \pi \sqrt{2} \sigma u^2 V \left(v \right)}{\nu_{\rm pas} \gamma^2} \frac{\partial f_0}{\partial r }\right] \Upsilon\left(z \right).
\end{equation}
Then, our kinetic equation reduces to
\begin{equation}
    \partial^2 \Upsilon / \partial z^2 - i z \Upsilon = -1.
\end{equation}
This has the particular solution of \cite{su1968collisional}
\begin{equation}
    \Upsilon_{SO} \left(z \right)  = \int_0^{\infty} d \tau e^{-iz \tau -\tau^3/3}.
\end{equation}
The homogeneous solution, subject to the constraint that it must vanish at $z \rightarrow \infty$, is 
\begin{equation}
    \Upsilon_{h} \left(z \right) = \Ai\left(ze^{i\pi/6}\right),
\end{equation}
where $\Ai$ is the Airy function. To find the overall solution, we apply the constraint that the plasma response must vanish at the trapped-passing boundary $z = -1/u$, giving
\begin{equation}
\label{eq:upsilon}
\Upsilon = \Upsilon_{SO} \left(z \right)  + \Upsilon_{\sqrt{\nu}} \left(z \right),
\end{equation}
where we have defined
\begin{equation}
    \Upsilon_{\sqrt{\nu}} \left(z \right) \equiv - \Upsilon_{SO} \left(-1/u \right) \Ai\left(z e^{i \pi/6}\right) / \Ai\left(-e^{i\pi/6}/u \right).
\end{equation}
This expression is plotted in Figure~\ref{fig:resresponse}, showing that $\gamma$ controls the proximity of the response to the freely passing location $k=0$ and $u$ controls the width of the resonance response. 

\section{Evaluation of flux} 
\label{sec:flux}
In this Section, we show how to use the alpha distribution perturbation $f_1$ derived in the previous Section to obtain the resulting alpha energy flux. This energy flux is the quantity of practical use in understanding and predicting tokamak discharges. In addition, we discuss key parts of the expression for this flux.
\subsection{Setup of flux expression}
The alpha particle energy flux of~\eqref{eq:hflux} is evaluated by recalling~\eqref{eq:d3v} to obtain
\begin{equation}
\Gamma_q \approx -\frac{M_\alpha }{2 R B_p \oint d \vartheta/ \vec{B} \cdot \nabla \vartheta} \int d \vartheta d \zeta dv d \lambda \frac{v^5 v_\parallel}{ \vec{B} \cdot \nabla \vartheta \left| v_\parallel \right|} f_1 \partial_\zeta A_\parallel.
\end{equation}
Using~\eqref{eq:apar},~\eqref{eq:fourierf1v2}, and~\eqref{eq:F1}, and the approximate expression $d\lambda \approx 4 \epsilon dk$, we can simplify this expression to
\begin{equation}
\Gamma_q = -\frac{4\sqrt{2} \pi^2 \epsilon M_\alpha n A_\parallel  }{R B_p \nu_{\rm pas}} \int_0^{v_0} dv \int_0^1 dk \frac{v^5 u^2 V\left(v \right)}{\gamma^2} \frac{\partial f_0}{\partial r} \Re\left[ \Upsilon \left(z \right) \right].
\end{equation}
Using the expression $ dk = - \left(u/\gamma \right)dz$, the pitch angle integral is performed for $u \ll 1$ by using the asymptotic forms in \cite{abramowitz1968handbook} to find
\begin{equation}
\label{eq:form}
   \int_0^1 dk \Re{ \left[\Upsilon\left(z \right) \right]} \approx \frac{u\pi}{\gamma} \left( 1 + \frac{u^{3/2}}{\sqrt{2} \pi} \right).  
\end{equation}
The first term in~\eqref{eq:form} is the resonant plateau contribution to the energy transport, while the $u^{3/2} \propto \nu_{\rm pas}^{1/2}$ term is the small correction due to the presence of the trapped-passing boundary. As important contributions to the speed integral are from the $v$ near the birth speed, we can use the approximate expression
\begin{equation}
    \frac{\partial f_0}{\partial r} \approx \frac{\partial n_\alpha /\partial r }{4 \pi v^3 \ln{\left(v_0/v_c\right)}},
\end{equation}
along with the speed dependent~\eqref{eq:udef} and the expressions in Table~\ref{tab:quants}, to write our energy flux as 
\begin{equation}
    \Gamma_q = -\frac{\sqrt{2} \pi^2 n A_\parallel^2 M_\alpha  q^2 \Omega_p}{\sqrt{\epsilon} B_0^2 \ln{\left(v_0/v_c\right)}} \frac{\partial n_\alpha}{\partial r} \int_0^{v_0} dv \frac{v}{\gamma}.
\end{equation}
Then, let us write
\begin{equation}
    \gamma = e^{2\Pi_0 v_0/v} 
\end{equation}
where we have defined
\begin{equation}
    \Pi_0  \equiv \frac{\sqrt{2 \epsilon} \Omega_p R \pi \sigma \left(nq-m \right) }{nv_0q},
\end{equation}
such that we can write
\begin{equation}
    \int_0^{v_0 } \frac{v}{\gamma} dv \approx \frac{v_0^2 e^{-2\Pi_0} }{2 + 2\Pi_0} \equiv \frac{v_0^2}{2} G \left(\Pi_0 \right),
\end{equation}
where $2\Pi_0 \rightarrow 2 \Pi_0 +2 $ is inserted to avoid singular behavior at $qn = m$ and recover the appropriate value as $\Pi_0 \rightarrow 0$.
Then, ignoring the small $u^{3/2}$ correction from the trapped-passing boundary, our energy flux is
\begin{equation}
\label{eq:heatflux3}
   \Gamma_q =  -\frac{\sqrt{2} \pi^2 n A_\parallel^2 M_\alpha  q^2 \Omega_p v_0^2 G\left(\Pi_0\right) } {2 \sqrt{\epsilon} B_0^2 \ln{\left(v_0/v_c\right)}} \frac{\partial n_\alpha}{\partial r}.
\end{equation}
From here, we can also define a diffusion coefficient:
\begin{equation}
\label{eq:D}
    D_{rp} = -\frac{2 \Gamma_q }{ \left(\partial n_\alpha/\partial r\right) M_\alpha v_0^2} = \frac{\sqrt{2} \pi^2 n A_\parallel^2  q^2 \Omega_p G\left(\Pi_0 \right)}{\sqrt{\epsilon} B_0^2 \ln{\left(v_0/v_c\right)}} .
\end{equation}
We next show how a phenomenological estimate of this diffusivity can be made for $\Pi_0 \sim 1$.
\subsection{Comparison to phenomenological estimates}
Let us compare~\eqref{eq:D} to an expression that can be arrived at via phenomenological methods. In particular, the resonant plateau diffusivity can be estimated as 
\begin{equation}
    D_{rp} = F_{\rm trap} V^2/\nu_{\rm eff},
\end{equation}
where $F_{trap}$ is the fraction of particles in resonance, $V$ is the radial drift, and $\nu_{eff}$ is the effective collision frequency. The width of the resonance in $k$ is given by 
\begin{equation}
u_k = \frac{u}{\gamma} = \left(\frac{\nu_{\rm pas}}{\gamma \omega_d}\right)^{1/3},
\end{equation}
where $u$ is defined in~\eqref{eq:udef} (The additional $\gamma$ relative to~\eqref{eq:udef} reflects the factor of $\gamma$ in~\eqref{eq:xdef}.) Then, the fraction of passing particles in resonance is
\begin{equation}
F_{\rm trap} = \sqrt{\epsilon} u_k.
\end{equation}
The radial drift is $V\left(v\right)$, defined in Table~\ref{tab:quants}.
The effective collision frequency is
\begin{equation}
    \nu_{\rm eff} = \frac{\nu_{\rm pas}}{u_k^2} = \left( \nu_{\rm pas}^{1/3} \gamma^{2/3} \omega_d^{2/3}\right).
\end{equation}
Then, the diffusivity can be estimated as
\begin{equation}
\label{eq:Drp}
    D_{rp} = \frac{\sqrt{\epsilon} V^2}{\gamma \omega_d} = \frac{ n q^2 \Omega_p  A_\parallel^2}{\gamma \sqrt{\epsilon} B_0^2},
\end{equation}
in qualitative agreement with~\eqref{eq:D}.

\subsection{Constraints on transport from quasilinear calculation}
To avoid significant transport, we require
\begin{multline}
    1 \gg \frac{D_{rp} \tau_s}{a^2} = \frac{\sqrt{2} n A_\parallel^2 \pi^2 q^2 \Omega_p \tau_s G\left(\Pi_0 \right)} {\tau_s \sqrt{\epsilon} B_0^2 \ln{\left(v_0/v_c\right)} a^2} 
    \\
    = 
    \left( \frac{B_\perp}{B} \right)^2 \frac{ \sqrt{2} \epsilon^{3/2} R^2 \pi^2 \Omega_p \tau_s G\left(\Pi_0 \right)}{a^2 n \ln\left(v_0/v_c\right)} ,
\end{multline}
where in the final equality we have used $B_\perp/B_0 = nA_\parallel/\left(R B_p\right)$.
Using values in Table~\ref{tab:equilib}, we calculate that $\tau_s = 0.2 \, \rm s$, $\Omega_p = 10^8 \, \rm s^{-1}$, $v_0/v_c \approx 2.3$, $\epsilon \approx 1/5$, and our expression becomes 
\begin{equation}
\label{eq:limit}
    \left(B_\perp/B_0\right)^2 \ll 10^{-9} / G\left(\Pi_0 \right).
\end{equation}
These results show that concern for NTM-driven alpha transport loss is well-founded as~\eqref{eq:limit} implies even a very small NTM amplitude can lead to substantial alpha particle energy loss. In the next Section we estimate when we expect our treatment begins to fail because the unperturbed axisymmetric tokamak alpha particle trajectories are perturbed by the NTM magnetic field.

\section{Conclusions and discussion}
\label{sec:concl}
We have evaluated the quasilinear transport of alpha particles caused by neoclassical tearing modes in a tokamak. We have shown that this transport can be significant even at small mode amplitudes, suggesting that it may influence the physics of next generation devices. 

A major limitation of our work is that it does not include the nonlinear physics which is in practice important to realistic neoclassical tearing modes which might be encountered in plasma experiments. Our quasilinear treatment will fail when nonlinear terms in the kinetic equation become comparable to linear ones, i.e., when
\begin{equation}
\label{eq:quasilin}
    \frac{ V \partial f_1/\partial r}{f_1 \nu_{\rm pas}/u_k^2 } \sim \frac{ V f_1/\left(u_k R\right)}{f_1 \nu_{\rm pas}/u_k^2 }\sim \frac{V}{\gamma^{1/3} \nu_{\rm pas}^{2/3} \omega_d^{1/3} R } \sim 1.
\end{equation}
This shows our results will fail at low collisionalities or high mode amplitudes, suggesting nonlinear work is necessary to fully understand the transport.  We suspect that some of the physics insight obtained in our work may nonetheless be important in understanding nonlinear transport. If the estimate of~\eqref{eq:quasilin} is used along with~\eqref{eq:Drp}, we obtain the condition for the NTM-driven alpha transport to remain small as our quasilinear transport treatment fails:
\begin{equation}
    \left(\nu_{\rm pas} /\gamma \omega_d \right)^{1/3} \ll \epsilon^{5/2} v_0^3/v_\lambda^3,
\end{equation}
implying there is a meaningful regime of validity for our results, but our quasilinear treatment will fail before the alpha energy transport loss becomes substantial. As a result, our resonant plateau transport estimate may be overly pessimistic. Indeed, we expect the NTM-perturbed alpha particle motion will diminish the resonant interaction and lead to some reduction in the transport provided stochastic behavior due to island overlap is avoided.

\section{Acknowledgements}
E.T. was supported by the W.M. Keck Foundation Fund at the Institute
for Advanced Study. Research at the Flatiron Institute is
supported by the Simons Foundation. This work was supported by the U.S. Department of Energy under contract number DE-FG02-91ER-54109. The United States Government retains a non-exclusive, paid-up, irrevocable, worldwide license to publish or reproduce the published form of this manuscript, or allow others to do so, for United States Government purposes.

\section{Declaration of interests}
The authors report no conflict of interest.

\appendix
\section{NTM description}
\label{sec:ntm}
In this Appendix, we present a heuristic description of tearing mode physics. The equilibrium magnetic field along the direction of the NTM wavevector is zero at the resonant surface and reverses direction on either side of it due to the shear in the magnetic field. Such a reversing field could possibly be unstable to a plasma instability known as the tearing mode \citep{furth1963finite,coppi1976resistive}.  Analysis of the tearing stability of a reversing magnetic field in non-toroidal slab geometry shows that the nonlinear growth rate is given by \cite{rutherford1973nonlinear}
\begin{equation}
\label{eq:dwdt}
\frac{dw}{dt} = \frac{\eta c^2}{4 \pi} \Delta' \left(w \right).
\end{equation}
Here, $w$ is the width of the tearing eigenmode, equivalent in our case to \citep{wesson2004tokamaks}    
\begin{equation}
    w = 4 \left[ \frac{q A_\parallel \left( \psi_s \right) }{ \partial_r q B_p} \right]^{1/2},
\end{equation}
and $\Delta'\left(w\right)$ is the tearing mode index that allows the narrow, inner tearing layer to be matched to the outer, global NTM eigenfunction of interest here as the alpha transport drive.

This instability is modified somewhat in tokamak geometry due to the naturally-driven bootstrap current present in tokamaks.  A heuristic derivation of this current follows, based directly on that typically available in pedagogical sources (for example, \cite{wesson2004tokamaks} and \cite{helander2005collisional}). The fraction of trapped particles in a tokamak is given by 
\begin{equation}
    f_{\rm trap} \sim \sqrt{\epsilon}, 
\end{equation}
where $\epsilon$ is as defined in~\eqref{eq:eps}. These trapped particles have a parallel velocity given approximately by 
\begin{equation}
    v_\parallel \sim \sqrt{\epsilon}v_{th};
\end{equation}
see for example, equation 5.22 of \cite{tolman2020drift}. The width of a banana orbit is given by
\begin{equation}
    w_b \sim \frac{B\sqrt{\epsilon} \rho}{B_p}.
\end{equation}
In a tokamak, a density gradient exists across the minor radius.  This means that the barely passing electrons will carry a current somewhat similar to a typical diamagnetic current in a plasma (note, however, that the current is in the parallel direction and enhances the Ohmic current). Considering the collisional exchange between electrons and ions and between trapped and passing populations gives that 
\begin{equation}
\label{eq:jb1}
    j_b \sim -\frac{\epsilon^{1/2}}{B_p} T \frac{dn}{dr}.
\end{equation}
A more precise derivation shows that temperature gradients also contribute to the bootstrap current, such that we can write 
\begin{equation}
\label{eq:jb}
    j_b \sim -\frac{\epsilon^{1/2}}{B_p} \frac{dp}{dr}.
\end{equation}
Notably, this bootstrap current depends on the gradient in pressure and is carried by the barely passing electrons as implied by the $\epsilon^{1/2}$ factor.  This means that if MHD activity creates an island in the magnetic field over which the background plasma pressure gradient is flattened, this will yield a change in the bootstrap current. This change is destabilizing to the tearing mode, such that the nonlinear growth rate~\eqref{eq:dwdt} is modified to become \cite{wesson2004tokamaks}
\begin{equation}
    \label{eq:dwdt2}
    \frac{dw}{dt} = \frac{\eta c^2}{4 \pi} \left[ \Delta' \left(w \right) + \frac{\alpha_{pq} \epsilon^{1/2} \beta_p}{w}\right],
\end{equation}
where 
\begin{equation}
\beta_p \equiv \frac{8 \pi  p }{B_p^2}
\end{equation}
and 
\begin{equation}
    \alpha_{pq} \equiv -\frac{8 p' q}{p q'},
\end{equation}
a value that is usually positive. This equation indicates that tearing mode instability can exist even in cases where $\Delta'<0$, and in fact suggests that instability should always exist for sufficiently small $w$. In fact, islands that are too small cannot flatten the pressure gradient effectively \citep{wesson2004tokamaks}. A seed magnetic island, perhaps due to a sawtooth or fishbone, can seed a perturbation in the bootstrap current, creating a magnetic island over which pressure gradients are significantly flattened \citep{buttery2000neoclassical,la2006neoclassical}. From~\eqref{eq:dwdt2} we can find the equation defining the  saturated neoclasssical tearing mode island width, $w_{\rm sat}$, as 
\begin{equation}
    \Delta' \left( w_{\rm sat} \right) = - \frac{\alpha_{pq} \sqrt{\epsilon} \beta_p}{ w_{\rm sat}}.
\end{equation}


\bibliographystyle{jpp}

\bibliography{jpp-instructions}

\end{document}